\documentstyle[aps,prl,twocolumn]{revtex}
\def\bra#1{{\langle#1|}}
\def\ket#1{{|#1\rangle}}
\def\tr{{\rm Tr}}

\begin{document}

\title{Distillation of GHZ states by \\
selective information manipulation}

\author{Oliver Cohen$^{1,2}$ and Todd A. Brun$^1$ \\
$^1$Physics Department, Carnegie Mellon University, \\
Pittsburgh, PA  15213  USA \\
$^2$Physics Department, Birkbeck College, University of London, \\
Malet Street, London  WC1E 7HX, England }

\maketitle

\begin{abstract}
Methods for distilling maximally entangled
tripartite (GHZ) states from arbitrary entangled
tripartite pure states are described. These
techniques work for virtually any input state.
Each technique has two stages which we
call primary and secondary distillation.
Primary distillation produces a GHZ
state with some probability, so that when
applied to an ensemble of systems, a
certain percentage is discarded.  Secondary
distillation produces further GHZs
from the discarded systems.
These protocols are developed with
the help of an approach to quantum 
information theory based on
{\it absolutely selective information}, which has
other potential applications.
\end{abstract}

In the rapidly developing field of quantum information, it is possible to
identify two main lines of investigation. On the one hand, it
addresses basic questions on the fundamental nature of information,
how it is embodied in quantum systems, how it can be quantified, and the
extent to which physical properties can be reduced to informational ones
\cite{QuantumInfo}. On the other hand are specific operational issues;
for example, how quantum information can be manipulated for
applications such as quantum computation and teleportation \cite{Shor}.

In this Letter we try to bring together these two strands by proposing a new
approach to the analysis of quantum information at a fundamental level,
which leads directly to an operational technique for distilling
maximally entangled tripartite (GHZ) states \cite{GHZ}, using local operations
and classical communication.  The three qubits of the system are assumed
to be physically separated and held by Alice, Bob, and Cara, respectively.
(Throughout this Letter we use
the term ``maximally entangled'' to refer to N-partite states that are
N-orthogonal: i.e., if the subsystems are two-dimensional such a state would
be ${\sqrt{1/2}}(\ket{0000\ldots} - \ket{1111\ldots})$.)

Central to our approach is the notion
of \textit{absolutely selective} information, which has a straightforward
interpretation in terms of classical information but can be
seen as a basic distinguishing feature between quantum systems and
their classical counterparts. We apply our approach to the specific problem
of distilling GHZ states from arbitrary entangled tripartite pure states.
We show that it is possible to distill,
with a certain probability, a GHZ state from virtually any entangled
tripartite pure state while retaining all three subsystems of the input
state. As far as we know this is the first protocol of this type to be
suggested. Our initial yield of GHZ states
is then supplemented by an additional yield which involves sacrificing some
subsystems.  In this Letter we outline our approach and
summarize our results. A more detailed exposition of the underlying
analysis will be presented elsewhere \cite{BrunCohen2}.

The distinction between ``selective'' and ``structural'' information was
addressed by Mackay \cite{Mackay} in the early days of classical information
theory.  Whereas structural information measures are based on an analysis of
the {\it form} of possible events, selective information refers to
{\it new} information gained from the occurrence of
a specific event.  For example, a signal might transmit a bit as one
of two different waveforms; the selective information would be one bit,
while the structural information, sufficient to describe all possible
waveform measurements, would be considerably more.
{\it Absolutely} selective information signifies data that are
irreducibly unpredictable, and hence genuinely new, in the sense that their
unpredictability cannot be explained by the observer's ignorance.
This type of information can arise only in a theory that is
fundamentally stochastic; hence it is commonplace
in quantum physics, but absent from classical physics.  For a pure state,
the minimum local absolutely selective information
(the minimum information generated by measuring one of the subsystems
with a free choice of measurement basis) is exactly the same
as the local entropy.

When considered as a quantitative measure, selective information is
closely related to fundamental measures in quantum information theory.
For example, the standard measure of entanglement for bipartite pure states
\cite{Bennett} is numerically equal to the minimum local absolutely
selective information. In a similar way, minimizing the absolutely
selective information can be used to develop measures of nonorthogonality
for quantum states \cite{Cohen2}. In this Letter we show that absolutely
selective information can be manipulated
by an appropriate measurement procedure, and apply this
to an operational problem.

The problem we address is to
transform a state $\ket{\psi_{123}}$
\begin{eqnarray}
\ket{\psi _{123}} &=& a \ket{000} + b \ket{001}
  + c \ket{010} +d \ket{011} \nonumber\\
&& + e \ket{100} + f \ket{101}
  + g \ket{110} +h \ket{111}
\label{psi123}
\end{eqnarray}
into the state
$\ket{\psi_{\rm GHZ}} = \sqrt{1/2} (\ket{000} -\ket{111})$,
with some probability.
Let us first consider the minimum absolutely selective
information $A_{i}$ associated with each of the three qubits in
$\ket{\psi_{123}}$:
\begin{equation}
A_i = - \left( p_i \log _2 p_i
  + \left( 1-p_i\right) \log_2 \left( 1-p_{i}\right) \right) 
\ \ \ \ \ i=1,2,3,
\end{equation}
where $p_i$ and $1-p_i$ are the eigenvalues of the reduced
density operator $\rho_i$ that describes system $i$ when the
other two subsystems are traced out,
\begin{equation}
\rho_i = \tr_{jk}\left( \ket{\psi_{ijk}} \bra{\psi_{ijk}} \right) .
\end{equation}
(Without loss of generality we adopt the convention that
$p_i \geq 1/2$.)

It can be shown \cite{Schlienz} that the \textit{maximal} value of
$\sum_{i=1}^3 A_{i}$
for any tripartite pure state occurs uniquely for the GHZ state (or any
local unitary transform of it), for which each of the $p_i$'s
is equal to 1/2 and $\sum_{i=1}^{3}A_{i} = 3$.  Our distillation procedure
makes use of this fact by decreasing each of the $p_i$'s in turn
(with some probability),
applying the procedure repeatedly, until all of the $p_i$'s are within
some tolerance of 1/2, at which point a GHZ state will necessarily have been
distilled (up to a local unitary transformation).

The technique for decreasing the $p_i$'s is similar to the ``Procrustes''
method of \cite{Bennett}.  We perform a
positive operator valued (POV) measurement consecutively on each subsystem.
To see how this works, let us carry out the procedure on
subsystem 1. If we consider subsystems 2 and 3 as a single composite
system, we can write the tripartite state as a Schmidt decomposition with
respect to subsystem 1 and the composite 2-3 system:
\begin{equation}
\ket{\psi _{123}} = \sqrt{p_1}\ket{+}_1 \ket{\phi^{+}}_{23} + 
  \sqrt{1-p_1} \ket{-}_1 \ket{\phi^{-}}_{23} .
\end{equation}
The minimum absolutely selective information for subsystem 1 
is then given by
\begin{equation}
A_1 = -\left( p_1 \log_2 p_1 +
\left( 1-p_1 \right) \log_2 \left( 1-p_1 \right) \right).
\end{equation}
By carrying out an appropriate POV measurement on this
subsystem, we can with some probability bring $A_1$ to
its maximal value of 1.  We introduce an ancilla qubit
``$a$'', which interacts unitarily with subsystem 1:
\begin{eqnarray}
\ket{+}_1\ket{0}_{a} &\longrightarrow&
  \alpha \ket{+}_1 \ket{0}_{a} 
  + \sqrt{1-\alpha^2}\ket{+}_{1} \ket1_{a} , \nonumber\\
\ket{-}_1 \ket{0}_{a} &\longrightarrow&
  \ket{-}_1 \ket{0}_{a} , \nonumber\\
\ket{+}_{1}\ket{1}_{a} &\longrightarrow&
  \sqrt{1-\alpha ^{2}}\ket{+}_1 \ket{0}_{a} - 
  \alpha \ket{+}_1 \ket{1}_{a} , \nonumber\\
\ket{-}_1 \ket1_{a} &\longrightarrow&
   \ket{-} _{1} \ket{1}_{a} .
\label{ancilla}
\end{eqnarray}
We then measure the state of the ancilla.
If we set $\alpha =\sqrt{(1-p_1)/p_1}$ and the starting
state of the ancilla to be $\ket{0}_{a}$, we will with
probability $2\left( 1-p_1 \right)$ find the ancilla in
state $\ket0_a$, which projects the system into
the state $\sqrt{1/2}\left( \ket{+}_1 
\ket{\phi^{+}}_{23} + \ket{-}_1 \ket{\phi^{-}}_{23}\right)$,
for which $A_1=1$.  With probability $2p_1-1$ we will measure state
$\ket1_a$, in which case the procedure fails.

In the Procrustes technique, this one step plus a local unitary
transformation suffices to distill EPR pairs
from arbitrary entangled bipartite states \cite{Bennett}.
In the tripartite case, we then repeat the procedure
on subsystems 2 and 3, which projects the
system into states for which $p_2=1/2, A_2=1$
and $p_3=1/2, A_3=1$, respectively.  However, if we simply
carry out a single POV measurement of this type on each of the three
subsystems in turn, the resulting tripartite state will not in general be a
GHZ state.  Each step of the process is nonunitary, since the
tripartite system can be discarded at each stage if the wrong result for the
POV measurement is obtained.  Whilst the $A_i$ are conserved by
unitary operations on the local subsystems,
they are not conserved in general for nonunitary
operations.  Hence, when we carry out a POV measurement on bit $i$ to project
the system into a state for which $p_i=1/2$ and $A_i=1$, this
will disrupt the values of $p$ and $A$ for the other two qubits.

Nevertheless, it transpires
that for most tripartite states, repeated application of this type of POV
measurement will steadily move the input state towards a GHZ state until it
gets arbitrarily close to it. (Exceptions will be identified later).  There
are a number of plausible ways to measure
``closeness'' to a GHZ state.  Three such measures are
\begin{mathletters}
\begin{eqnarray}
D_p &\equiv& \sum\limits_{i=1}^{3} p_i - 3/2 , \label{D_p} \\
D_S &\equiv& 3 - \sum\limits_{i=1}^{3} A_i , \label{D_S} \\
D_2 &\equiv& 3/4 - \sum\limits_{i=1}^{3} p_i\left( 1-p_i\right) .
\label{D_2}
\end{eqnarray}
\end{mathletters}
We introduce this last quantity because it is more tractable
analytically than $D_p$ and $D_S$, being a simple function of
the coefficients of $\ket{\psi_{123}}$.
%For example, for the starting state given by eq. (\ref{psi123}),
%we find that $p_{1}\left( 1-p_{1}\right) =\ket{af-be\right| ^{2}+\left|
%ag-ce\right| ^{2}+\ket{ah-de\right| ^{2}+\ket{bg-cf\right| ^{2}+\left|
%bh-df\right| ^{2}+\ket{ch-dg\right| ^{2}$, and similar expressions can be
%derived for $p_2\left( 1-p_2\right)$ and $p_3\left( 1-p_3\right)$.

Numerical analysis of this process on randomly chosen
starting states shows that for
a large fraction of these states, $D_p$ approaches zero
to an accuracy of $10^{-3}$ after just two complete iterations
(i.e. two POV measurements performed on each of the three subsystems),
while virtually all do so within four iterations.
Interestingly, we find that in every case examined (aside from the
exceptions given below), $D_p$ decreases monotonically toward zero
with each step of the procedure, whereas $D_S$ and 
$D_2$ can fluctuate, though of course their general trend is downward.

The results presented so far are supported only by numerical analysis.
However, there is a closely related procedure for which we have derived
an analytical proof of efficacy for virtually any input state.  In this
second method, instead of reducing each probability $p_i$ to $1/2$ in turn
using POV measurements, the probabilities are reduced by a small
amount $\epsilon$, so that with each step the state changes infinitesimally
in the limit $\epsilon\rightarrow0$.  The proof follows fairly
straightforwardly by deriving the changes in the state coefficients
from the procedure, then using these to derive the change in $D_2$.
By changing to the Schmidt basis for all three bits, one can, with some
effort, show that $p_i(1-p_i)$ can never decrease for $i=1,2,3$,
and will only remain unchanged for certain very special initial states
detailed below.  This monotonicity implies monotonicity for
$D_2$, $D_p$, and $D_S$, so all of these quantities diminish steadily
as the state approaches a GHZ state.
This infinitesimal method would be quite challenging experimentally,
but it is analytically interesting due to its relative tractability.

The protocols described so far correspond to what we call ``primary''
distillation. They will give a specific yield (i.e., surviving percentage)
of GHZ states if a collection
of systems is supplied in a given input state. This yield can be
straightforwardly calculated for the large-step procedure; after each
POV measurement on the $i$th subsystem a proportion $2(1-p_i)$
of the systems are retained.
The yield of GHZ states for the primary distillation process averaged
approximately $9.2\%$ for the evenly-distributed
sample of input states we analyzed, but this will clearly depend strongly
on the initial distribution.

Average yields for the infinitesimal procedure were $9.7\%$.
The chance of the procedure failing on any given step
is quite small, but over many steps the number of failures mounts.
The difference between the infinitesimal and big-step procedure is
interesting when contrasted with the bipartite Procrustes technique.
In the bipartite case, there is no advantage to using small steps
over a single large step; the yields are the same in both cases.  Clearly
in the more elaborate tripartite procedure there is a difference.

This yield can be greatly enhanced by a process of {\it secondary
distillation}, which makes use of those systems discarded during
primary distillation.  When we carry out the initial POV
measurement on subsystem 1 for the input state $\ket{\psi_{123}}$
given by eq. (\ref{psi123}), with probability $2p_1-1$
we will fail to obtain the desired result.
However, this failure will leave the discarded system in
the state $\ket{+}_1\ket{\phi^{+}}_{23}$,
where $\ket{\phi ^{+}} _{23}$\ is in general an entangled
bipartite state of subsystems 2 and 3. Similarly, failures at later steps of
the primary distillation process can yield entangled bipartite states of
subsystems 1 and 2 and of subsystems 1 and 3. Thus, when the primary
distillation procedure has been completed on a collection of systems in a
given input state, we will have an additional residue of entangled bipartite
states of subsystems 1 and 2, 1 and 3, and 2 and 3. These entangled pairs
can be distilled to EPR pairs by standard techniques \cite{Bennett}, and the
resulting EPR pairs can be used to prepare further GHZ triplets. 

This is quite similar to the method of \cite{DurCirac},
where GHZ states were produced by first distilling EPR pairs between
Alice, Bob, and Cara, and then using two pairs to produce a GHZ triplet.
(For example, if Alice shares one EPR pair with Bob 
and another with Cara, she can distribute
a GHZ state by preparing it locally and then teleporting the states of two
of the subsystems to Bob and Cara with the help of the two EPR pairs.)
If, when primary distillation is completed, we
produce $N_{23}$ EPR pairs of subsystems 2 and 3, 
$N_{31}$ EPR pairs of subsystems 3 and 1, and $N_{12}$
EPR pairs of subsystems 1 and 2 from the discarded systems, we will be
able to distill a further $(N_{23}+N_{31}+N_{12})/2$
GHZ triplets (in the case where none of the $N$s\ is greater than the
sum of the other two), or $(N_{jk}+N_{ki})$ GHZ triplets
(if $N_{ij}>N_{jk}+N_{ki}$).
Numerical analysis indicates that the average yield for secondary
distillation of GHZ states, for the random sample considered, is
approximately $27.5\%$ giving a total yield of about $36.7\%$.
The infinitesimal technique does even better, giving a secondary yield
of $29.3\%$ for a total yield of $39\%$.

Since the bulk of this yield comes from the production of EPR pairs,
one might reasonably ask how these methods compare to simply producing
EPR pairs (with no primary distillation), and then using these pairs to
produce GHZ triplets directly \cite{DurCirac}.  EPR pairs are produced
by measuring one of the subsystems in such a way as to maximize the
pair-wise entanglement between the other two bits, and then distilling
perfect EPR pairs from the resulting states.
For the same random sample of states,
this technique produces an average yield of $31.5\%$, lower than
either of the other two techniques, and not much higher than the
secondary yield alone.  This does not, of course, prove that it is
worse for every initial state.  However, the closer the initial state
is to a GHZ (using any of our distance measures (\ref{D_p}--\ref{D_2})),
the better these distillation procedures perform, while producing
GHZs from EPR pairs has a maximum yield of $50\%$.  (It may be possible
to increase this yield asymptotically; but if, so the secondary yield
in our protocols will also be increased, so we do not expect this
to change our conclusions substantially.)

For some special cases these protocols will not work as
described. If the original input state is not three-party entangled, the
protocol will fail completely; that is, if the original state can be written
as $\ket{\chi}_i\ket{\zeta }_{jk}$, no three-party
entanglement will be distillable by either primary or secondary
distillation.  There is
another set of states for which primary distillation fails, but which can
still produce GHZ states by secondary distillation.  This set consists
of tripartite input states with just three components, where each component
is biorthogonal (but not triorthogonal) to the other two, and local
unitary transforms of such states.
For example, the
state $\ket{\psi_{\rm tr}} =b \ket{001} + c \ket{010} + e\ket{100}$
is of this type. We call such states
``triple'' states; all have $D_p\ge 1/2$, though a substantial
subset has $D_p=1/2$ exactly.  Both forms of the
primary distillation process take triple states to triple states,
so that the GHZ state will never be produced. The large step
procedure causes all triple states to converge
to the state 
$\ket{\psi_{GM}} =
  \sqrt{1/2} \ket{001} + \sqrt{(\sqrt{5}-1)/4}\ket{010} 
  + \sqrt{( 3-\sqrt{5})/4}\ket{100}$, 
at which point any further steps
will simply result in a cyclic shuffling of the component amplitudes.
We call this attractor state the ``golden mean'' state because of the
appearance of the golden mean in the amplitudes.  The infinitesimal
procedure leaves all triple states with $D_p=1/2$ unchanged.
This procedure can also cause certain other states with
$D_p>1/2$ to converge to triple states rather than GHZ states, though 
most do not.  The large step procedure
may also take some states to triple states.  Although triple states do
not yield any GHZ states by primary distillation, they can of course produce
them by secondary distillation.

What is more, it is possible to move
off of a triple state (with some probability) by performing a POV measurement
in a basis other than the Schmidt basis.  That is, instead of using
the basis $\ket{+}_1$ and $\ket{-}_1$ in the transformation (\ref{ancilla}),
one uses a different basis, such as $(\ket{+}_1 \pm \ket{-}_1)/\sqrt2$.
Setting $\alpha$ to a reasonable value (such as $\alpha=\sqrt{1/2}$)
will then take triple states to non-triple states.  In
principle, the state might then re-converge to a triple state, but this is
quite unlikely.

We have shown that manipulation of absolutely selective information can be
used to distill maximally entangled tripartite states from arbitrary
tripartite entangled pure states.
This method will not work for systems with four or
more subsystems, since in these $p_i=1/2$ does not uniquely determine the
maximally entangled state.  It may be that related techniques might
succeed, however, if it is possible to manipulate other locally
unitarily invariant parameters by local POVs and classical communication.
Even in the 3 qubit case, however, the procedure we have
described is surely not optimal.  The optimal distillation technique
is not known, but would almost certainly make use of joint manipulations
on many copies of the input state.  Nor is an asymptotically reversible
distillation technique known for GHZ states \cite{Bennett2}.
It would be interesting to compare the
yields of these two hypothetical techniques.  In the bipartite case
they are the same, but this need not be so in the tripartite case.
Indeed, if the reversible GHZ distillation technique produced any
extra two-party entanglement, one would generally expect to be able
to produce further GHZ states by an irreversible secondary distillation
stage.  This suggests that the algorithm giving the optimal yield of
GHZs will probably not be reversible.
It would also be interesting to compare our yield of GHZ states to some
standard measure of tripartite entanglement.  Lacking such a measure,
however, the best that can be done is to compare different distillation
techniques to each other.

There may be a number of other problems in quantum information
theory which are amenable to an approach focusing on the absolutely
selective information content of quantum systems.  For example, work in
progress suggests than such an approach can be useful in the analysis of
nonorthogonality.  Since selective information is a classical concept,
this approach also provides a valuable link between classical and
quantum information.

We are grateful to Bob Griffiths and Chris Fuchs for helpful discussions.
This research was supported by NSF Grant No. PHY-9900755.


\begin{thebibliography}{99}

\bibitem{QuantumInfo} B. Schumacher, Phys. Rev. A 51, 2738 (1995);
S. Hill and W.K. Wootters,
Phys. Rev. Lett. 78, 5022 (1997); O. Cohen, Phys. Rev. Lett. 80, 2493
(1998); D.P. DiVincenzo and D. Loss, Superlattices and Microstructures 23,
419 (1998).

\bibitem{Shor} P.W. Shor, in \textit{Proceedings of the 35th Annual
Symposium on the
Foundations of Computer Science, Santa Fe, 1994}, edited by S. Goldwasser
(IEEE Computer Society Press, Los Alamitos, CA, 1994), p.124.

\bibitem{GHZ} D.M. Greenberger, M.A. Horne, and A. Zeilinger, in \textit{Bell's
Theorem, Quantum Theory, and Conceptions of the Universe}, edited by M.
Kafatos (Kluwer, Dordrecht, 1989), p.69.

\bibitem{BrunCohen2} T.A. Brun and O. Cohen, in preparation.

\bibitem{Mackay} D.M. Mackay,
\textit{Information, Mechanism, and Meaning} (MIT Press,
Cambridge, MA, 1969).

\bibitem{Bennett} C.H. Bennett, H.J. Bernstein, S. Popescu
and B. Schumacher, Phys. Rev. A 53, 2046 (1996).

\bibitem{Cohen2} O. Cohen, Phys. Rev. A 60, 4349 (1999).

\bibitem{Schlienz} J. Schlienz and G. Mahler, Phys. Lett. A 224, 39 (1996).

\bibitem{DurCirac} W. D\"ur and J.I. Cirac, quant-ph/9911044.

\bibitem{Bennett2} C.H. Bennett, S. Popescu, D. Rohrlich, J.A. Smolin
and A.V. Thapliyal, quant-ph/9908073.

\end{thebibliography}
\end{document}